\documentclass[preprint,3p,times,onecolumn]{elsarticle} %

\usepackage{graphicx}
\usepackage{subfig}    				   
\usepackage{epstopdf}

\usepackage{amssymb}
\usepackage{amsmath}

\usepackage[utf8]{inputenc}

\usepackage[displaymath]{lineno}
\modulolinenumbers[5]

\journal{arXiv}

\begin{document}

\begin{frontmatter}

\title{Multi-turn injection into a heavy-ion synchrotron in the presence of space charge}

\author[label1]{S. Appel}
\author[label1,label2]{O. Boine-Frankenheim}

\address[label1]{GSI, Helmholtzzentrum f\"ur Schwerionenforschung GmbH, Planckstr. 1, 64291 Darmstadt, Germany}
\address[label2]{Technische Universit\"at Darmstadt, Schlossgartenstra\ss e 8, 64289 Darmstadt, Germany}

\begin{abstract}
For heavy-ion synchrotrons an efficient Multi-Turn Injection (MTI) from the injector linac is crucial in order to reach the specified currents using the available machine acceptance. The beam loss during the MTI must not exceed the limits determined by machine protection and by the vacuum requirements. Especially for low energy and intermediate charge state ions, the beam loss at the injection septum can cause a degradation of the vacuum and a corresponding reduction of the beam lifetime. In order to optimize the injection of intense beams a very detailed simulation model was developed. Besides the closed orbit bump, lattice errors, the position of the septum and other aperture limiting components the transverse space charge force is included self-consistently. The space charge force causes a characteristic shift of the optimum tunes and a smoothing of the phase space density.  
\end{abstract}

\begin{keyword}
Synchrotron \sep multi-turn injection \sep space charge \sep simulation

\end{keyword}

\end{frontmatter}


\section{Introduction}\label{intro}
In heavy-ion synchrotrons the Multi-Turn-Injection (MTI) should fill up the available transverse acceptance, avoiding the already occupied phase space area~\cite{Gardner1991, wille2000physics, prior_mtj}. In order to increase the space charge limit, heavy-ion synchrotrons are operated with intermediate charge state ions. Therefore stripping injection is not an option and the MTI has to respect Liouville's theorem for the chosen charge state. The present study focuses on the MTI into the GSI heavy-ion synchrotron SIS18~\cite{parameterlisteSIS}. The SIS18 is presently being upgraded in order to increase the beam intensities for the FAIR (Facility for Antiproton and Ion Research) project~\cite{Spiller2008}. For the reference U$^{28+}$ ions the SIS18 intensities at injection energy (11.4 MeV/u) should reach the limit determined by transverse space charge and error resonances (space charge limit).
Heavy-ion ion beams are injected from the GSI UNILAC and are stacked along the horizontal plane using multi-turn injection ~\cite{parameterlisteSIS}. The beam loss should be well below 30\% to avoid an increase of the dynamic vacuum pressure as well as activation and damage. The transverse beam size of the stacked beam should be within the machine acceptance ($\epsilon_x=150$ mm mrad, $\epsilon_y=50$ mm mrad). Due to the elliptic beam pipe the vertical machine acceptance is much smaller than the horizontal. Also the resulting rms momentum spread ($\Delta p/p\leq10^{-3}$) has to remain below a maximum value determined by the available rf bucket area for fast ramping. In the SIS18 the microbunches injected from the UNILAC with 36 MHz debunch, filament and form a coasting beam within a few turns. During debunching the longitudinal space charge force of the intense microbunches leads to an increases of the final momentum spread~\cite{Appel2012}. In this study we will first discuss optimum multi-turn injection schemes for low intensities. Afterwards the impact of transverse space charge on the optimum MTI settings and the resulting phase space distribution is discussed.  

\section{Multi-turn injection model}\label{Multi_turn_injection_model}
%
\begin{figure}
\centering	
\includegraphics[scale=0.25]{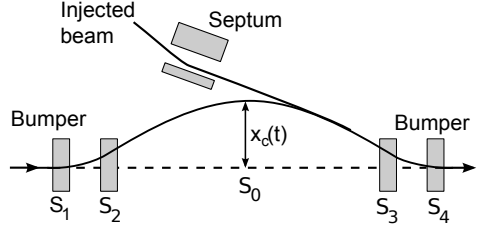}
\caption{The septum and bumper magnets\label{bumper}}
\end{figure}
\begin{figure}[htb]
\centering
\includegraphics[scale=0.4]{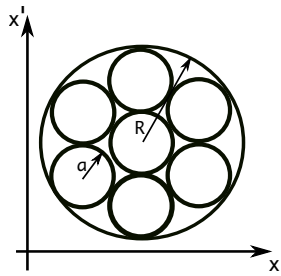}
\caption{Optimal packing of seven identical beamlets of radius $a$ into a machine acceptance with radius $R$.}
\label{packing}       
\end{figure}

In the SIS18 the injected beamlets are stacked in the horizontal phase space until the machine acceptance is reached. To fulfill Liouville's theorem, four bumper magnets create a time variable closed orbit bump such that the electrostatic injection septum deflects the next incoming beamlet into free horizontal phase space and close to the formerly injected beamlets. Fig. \ref{bumper} indicates the position of the four bumper magnets along the circumference. If we require that the four bumper magnets produce no closed orbit distortion outside of the injection region and at the injection position $s_0$, then the amplitude $x_0$ as well as the angle $x_0'$ of the closed orbit are the degrees of freedom. The resulting angular kicks produced by the bumper magnets are~\cite{Gardner1991,wille2000physics}

\begin{align}\label{phi}
		\varphi_1 &= \frac{d_{02}x_0-b_{02}x'_0}{b_{21}}\ ,& \varphi_2 &=\frac{-d_{01}x_0+b_{01}x'_0}{b_{21}}, &&\\
	\varphi_3 &= \frac{-b_{41}\varphi_{1}-b_{42}\varphi_2}{b_{43}}\ ,& \varphi_4 &= \frac{b_{31}\varphi_{1}+b_{32}\varphi_2}{b_{43}}
 &&\label{phi2}
\end{align}	
with
\begin{align}
	b_{lk} &= \sqrt{\beta_l\beta_k}\sin(\phi_l-\phi_k), &&\\
	d_{lk} &= \sqrt{\frac{\beta_k}{\beta_l}}\sin(\phi_l-\phi_k)-\frac{\alpha_l}{\beta_l}b_{lk}. &&\label{phi3}
\end{align}
Here $\alpha_l$ and $\beta_l$ are the horizontal lattice parameters and $\phi_l$ is the phase advance at the point $s_l$. \\ 
To achieve high beam intensities the injected beamlets should be packed as compact as possible. In normalized phase space coordinates the injected beamlets as well as the beam pipe are approximately circular, therefore the MTI packing problem is similar to the packing of ropes and cables. Assuming that $n_{eff}$ beamlets with radius $a$ are packed into a given machine acceptance (or container) with radius $R$ like shown in Fig. \ref{packing} (also named hexagonal packing) and imagine hexagons which circumscribe those beamlets. Then the dilution is defined as the area of the container divided by the area of these hexagons~\cite{Kravitz1967} 
\begin{equation}\label{efficiency}
	 d = \frac{2\sqrt{3}}{\pi}\frac{R^2}{a^2}\frac{1}{n_{eff}}.
\end{equation}
The hexagonal packing of circles onto an annular ring leads to a shell structure and the dilution will be larger than~one. In~\cite{Kravitz1967} it is shown that for $n_{eff}=7$ and $n_{eff}=19$ the smallest dilution is achieved, namely $d_7=1.16$ and $d_{19}=1.12$.  \\ 
The incoming beamlets will have a linear $x_i$ and an angular $x'_i$ displacement with respect to the closed orbit and will therefore undergo betatron oscillations. One turn later the injected beamlets will pass the injection point again, but due to the betatron oscillation will avoid the septum. If the orbit is not reduced fast enough, the beamlets will hit the inner side of the septum after $n$ revolution turns, depending on the betatron oscillation tune and get lost. Therefore the beam loss has its maximum located at the resonance condition
\begin{equation}\label{resonance_condition}
	Q_x n =integer
\end{equation}
where $Q_x$ is the horizontal tune. During injection the number of injected turns $n_i$ and the revolution turns $n$ are equal, after injection this condition does not hold any longer because $n_i$ remains constant. If $\eta$ characterizes the ratio between the lost and injected particles it follows	
\begin{equation}\label{neff}
	n_{eff} = n_i(1-\eta).
\end{equation}
For a loss free injection $\eta$ is zero and the effectively accumulated beamlets are equal to the number of injected turns. The quality of the MTI is determined by the number of accumulated beamlets and the associated loss. For a given machine acceptance a high MTI quality implies a small dilution factor. \\
The beamlets injected at latter turns will have a larger betatron amplitude as the orbit bump is reduced. This leads to the formation of 
\begin{equation}
	m \sim 
	\begin{cases}
	 Q_f^{-1}, & \text{if } 0 <  Q_f \le 0.5 \\
	 (1-Q_f)^{-1}, & \text{if }0.5 < Q_f \le 1 
	\end{cases}	
\end{equation}
arms determined by the fractional tune $Q_f$. In our later considerations we will assume fractional tunes between $0.0$ and $0.5$. \\
To achive small dilution factors one would choose a fractional tune, which leads to a large number of arms. However, one has to consider the restriction given by the geometry and by the machine resonances. If the first beamlet is injected exactly on the closed orbit; one can place at maximum six beamlets with the same size around this beamlet as shown in Fig.~\ref{packing}. To avoid beam loss due to the already occupied phase space area, the fractional tune should not be smaller than $1/6$. But then one has to accept a larger dilution factor for the next shell than given by the geometry optimum (six instead of twelve). This could be improved by ramping the tune using pulsed quadrupoles. However, this option will not be considered in the present study.
\begin{figure}[htb]
\centering
\includegraphics[scale=0.2]{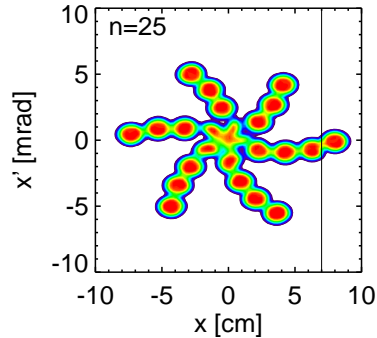}
\caption{Snapshot of the MTI from a simulation showing the shell structure in normalized phase space.}
\label{fig:ps_nosc}       
\end{figure}
\\
Fig. \ref{fig:ps_nosc} shows a snapshot of a MTI simulation in normalized horizontal phase space for a fractional tune of $Q_f=0.17$ resulting from the present arguments as optimal tune for the injection. The linear reduction of the bump per turn
\begin{equation}\label{bump_fall}
	\Delta x \approx Q_f (2a + d_s)
\end{equation}
in the simulation has been chosen such, that the incoming beamlets miss the inner side of the septum with their edges. The septum  thickness $d_s$ can be neglected for the SIS18 injection since the beamlet radius is in the range of $5$ and $10$~mm and the septum thickniss is $0.1$ mm. The vertical solid line in the figure indicates the septum. The newly injected beamlet is located to the right of the septum and will then oscillate towards the stacked beamlets to the left of the septum. The formation of six arms is clearly visible as well as the resulting shell structure. In the simulation the decoherence due to the chromaticity $\xi$ has been supressed to demonstrate more clearly the influence of the coherent betatron oscillations of the injected beamlets. The decoherence would fill the voids between the injected beamlets after many turns. \\
The MTI quality can be further optimised by adjusting the start value of the closed orbit bump, its ramping curve and the position of the incoming beamlet. For an optimal filling of the phase space (e.g. small dilution factor)  Ref.~\cite{prior_mtj} propose to map the injected emittances onto upright ellipses   
\begin{equation} \label{ratio_prior} 
	\frac{\alpha_0}{\beta_0}=-\frac{x'_i-x'_0}{x_i-x_0}
\end{equation}
and mismatch the lattice function 
\begin{equation}
	\frac{\beta_0}{\beta_i}=\left(\frac{\epsilon}{\epsilon_i}\right)^{1/3}
\end{equation}
in order to adapt the curvature of the incoming beamlet to the ring acceptance curvature. The subscript $i$ represents the parameter for the incoming beamlet and the subscript $0$ for the ring. $\epsilon$ is the available ring acceptance and $\epsilon_i$ is the emittance of the injected beamlet. The beam center of the incoming beamlet should approximately placed as such, that the edge of the incoming beamlet touches the septum at the outside (see Fig.~\ref{fig:ps_nosc})
\begin{equation}\label{pos}
	x_i \approx a + x_s + d_s + x_{st}
\end{equation}
with $x_s$ as the septum position and $x_{st}$ as the possible steering offset. The angular center position of the incoming beamlet is limited by the septum deflection angle $x'_s$ and the possible steering of the linac beam $x'_{st}$
\begin{equation}\label{ang}
	x'_i \approx x'_s + x'_{st}.
\end{equation}
For the orbit bump reduction also other ramps are proposed besides a linear ramp to better adapt the orbit bump to the free phase space regions. An exponential ramp is given through
\begin{equation}
	x(n) = x_0 \exp(-(2a+d_s)\tau n)
\end{equation}
for the orbit bump variation per turn. By varing $\tau$ one can possibly find an efficient ramping curve which provides an optimal filling of the phase space accompanied by low beam loss. To further improve the MTI quality also the steering parameters $x_{st}$ and $x'_{st}$ should be varied as long the mapping condition into upright ellipses is fulfilled. 

\section{MTI simulation model}

The MTI model described in the previous section has been implemented in the particle tracking codes PATRIC (see e.g. \cite{Boine-Frankenheim2008a}) and pyORBIT (see e.g \cite{Cousineaua, Shishlo2006}). Both codes use space charge solvers together with a Particle-In-Cell (PIC) numerical scheme. The simulation model includes the closed orbit bump, lattice errors and the position of the septum. To consider loss on machine acceptance we placed an acceptance collimator vis-$\grave{a}$-vis to the septum. The simulation model also contains the possibility to obtain the closed orbit bump from the SIS18 control program approximation. 
For technical reasons the present SIS18 control program uses an approximation for the calculation of the angular kick from the bumper magnets, since the evaluation of the horizontal lattice parameters as well as the phase advance at the bumper magnet position and at the injection point of a given horizontal tune are not possible. To find a function depence on horizontal tune for a fixed amplitude ($x_0=10$ mm) and angle ($x_0'=1$ mrad) the angular kicks for several horizontal tunes given by Eq. \ref{phi} were calculated and quadratic functions were fitted on these results. By normalizing this function with the bump amplitude of $x_0=10$ mm mrad one can adjust the four functions on each bump amplitude by multiplying them with the desired bump amplitude~\cite{Ondreka}. The new control system does not have such limitations and therefore this bump calculation will be used only for the comparison with current perform SIS18 experiments.\\ 
In PATRIC the Poisson equation is solved on a static 2D transverse grid together with an FFT solver and momentum kicks corresponding to the local space charge field strength are applied. For the macro-particle tracking PATRIC uses the linear transfer matrix computed by the lattice program MADX. \\
pyORBIT is a Python/C++ implementation of the ORBIT (Objective Ring Beam Injection and Tracking) code. The code is freely  available at \cite{Cousineaua}. In pyORBIT the transverse space charge potential for the particle distribution is evaluated on a dynamic 2D grid using an FFT solver and the kicks on the macro-particles are obtained from second order interpolation~\cite{Shishlo2006}. In pyORBIT the particles are tracked with a TEAPOT (Thin-Element Accelerator Program for Optics and Tracking~\cite{Schachinger1987}) method. pyORBIT allows to build the TEAPOT lattice from a MAD input file by analyzing the input file. We implemented a python routine to include the model for the closed orbit bump.

\section{Comparison of MTI simulations and experiment}
\begin{figure}
\centering
\includegraphics[scale=0.25]{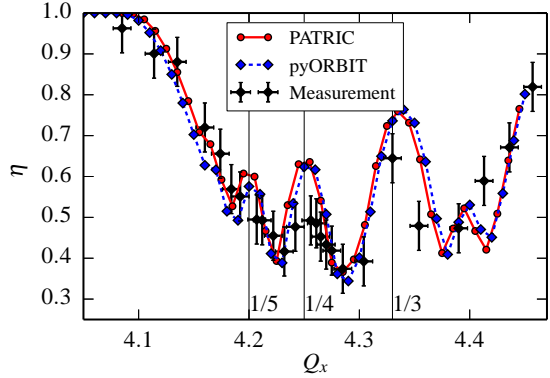}
\caption{Simulated and measured MTI loss as a function of the horizontal tune. The loss maxima given by Eq. \ref{resonance_condition} are indicated as vertical lines. The measurement results were provided by~\cite{Hayek2013}.}
\label{measurement}       
\end{figure}
\begin{figure}[htb]
\centering
\includegraphics[scale=0.25]{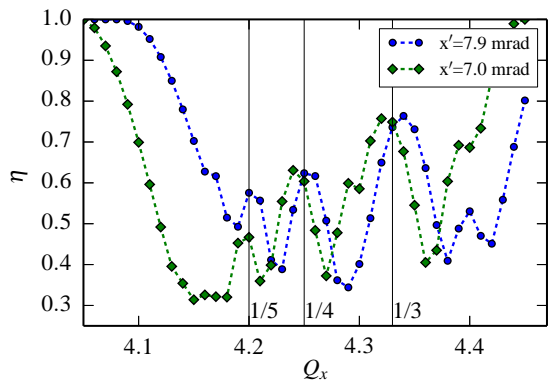}
\caption{MTI loss as a function of the horizontal tune for two different beamlet slopes.}
\label{angle}        
\end{figure}

Accurate predictions of the MTI efficiency require a careful validation of the machine model with experiments. The MTI efficiency depends on various machine and beam parameters. Often, some important MTI parameters are not precisely known or can change from cycle to cycle. Therefore, an accurate measurement of the horizontal tune is most importent. In the SIS18 the machine tune is obtained with very high precision from the transverse Schottky spectrum. During a dedicated MTI experiment with $U^{28+}$ ions \cite{Hayek2013}, besides the tunes also the injected emittance was measured ($\epsilon_i=6.5\pm 0.7$ mm mrad) in the transfer channel to the SIS18 a few meters before the injection point. The comparison between measurements and simulations is shown in Fig. \ref{measurement} for low currents.
The measurements as well as the simulations with both codes show beam loss maxima located at the resonant tunes. In the simulations the machine injection settings have been used (orbit amplitude of $x_0$ = 85 mm, bump reducing of $\Delta x_0$ = 2.5 mm per turn and the number of injected turns of $n_i=21$). In the SIS18 it is not possible to measure the slope and the position of the incoming beamlet at the septum. The shown agreement between measurements and simulations was obtained by setting the center of the injected beamlets to $x_i$ = 90 mm (approximately the position of septum plus twice the beamlet radius) and its slope to $x'_i$ = 7.9 mrad (the range of the linac steering angle is 1-3 mrad) in the simulations. \\
The experiment provided by~\cite{Hayek2013} indicate that the fraction tune of $Q_f=0.17$ is not always optimal for an efficiently injection into the SIS18. This discrepency can be explained by a mismatch of the beamlet slope in the experiment. In the simulation an even lower beam loss around $Q_f=0.17$ could be achieved by changing the beamlet slope. As shown in Fig. \ref{angle} for a beamlet slope of $x'_0=7.0$ mrad for tunes around 4.17 the loss is reduced by $30\%$. Also a shift of the maxima is observable. In other dedicated MTI experiments a lower loss for tunes around 4.17 could be obained~\cite{Hayek2013}.

\section{Improvement of the SIS18 MTI quality}
\begin{figure}
\centering	
\includegraphics[scale=0.275]{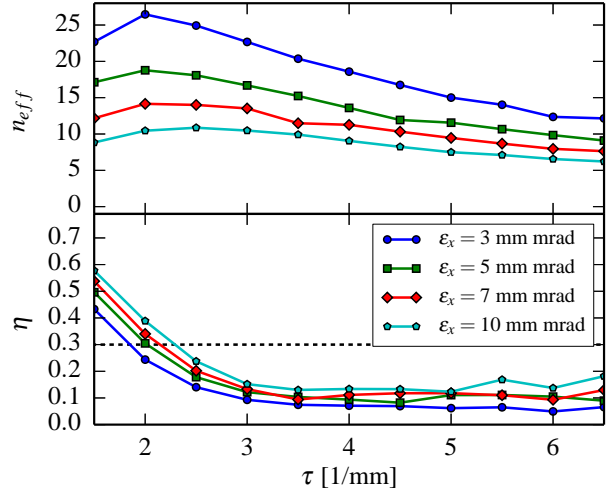}
\caption{The number of effectively stored turns and the beam loss as a function of the ramping rate $\tau$ for four different emittances. \label{fig:exp}} 
\end{figure}

In order to find an efficient ramping curve we performed several simulations by varying the parameter $\tau$ for different initial emittances (show in Fig.~\ref{fig:exp}). The chosen emittance values are in the range of the expected emittances for the planned UNILAC injector upgrade. In the simulation the injection is stopped if the rms emittance of the stacked beam exceeds the machine acceptance. This automatically maximizes the number of effectively accumulated beamlets for a given initial emittance. The chosen injection stop condition results in a different number of injected turns $n_i$ in each simulation. To take into account any possible subsequent beam loss the simulation continues for 200 turns. As indicated in Fig.~\ref{fig:exp} $\tau=3$ gives a good compromise between a compact packing of the beamlets and loss. For all emittances a MTI qualtiy with $\eta\approx0.15$ loss and dilution factor $d\approx2$ could be achieved. If higher loss during injection is acceptable a slower reduction (e.g. a smaller $\tau$) could be selected to reach a larger number of efficiently accumulated beamlets. The speed of the orbit reduction is reflected in the dilution factor. For a slower reduction the dilution could be reduced to $d=1.5$ at the cost of a larger loss. The simulation results indicate that for smaller emittances the dilution is slightly larger then for larger emittances. The achieved dilution factors are similar to the dilution factors of $d=1.4-2$ presented by \cite{Chao1999} for a conventional one plane multi-turn injection. \\
For the SIS18 an optimum steering offset of the incoming beamlet was found for $x_{st}=-0.2a$ and $x'_{st}=0.5$~mrad as well as $x'_i-x'_0=0.25$~mrad and $x_i-x_0=2.8$~mm. The ratio of the lattice parameters at the injection point in the SIS18 for the horizontal tune of $Q_f=4.17$ is $\frac{\alpha_0}{\beta_0}=-0.09$ and the SIS18 septum deflection angle is $x'_s=6$~mrad.

\section{Space charge effects}
The PATRIC and pyORBIT codes were used to study the effects of the transverse space charge force on the multi-turn injection. The incoherent tune shift for a round, incoming beamlet with a homogeneous density is \cite{schindl_space_charge}
\begin{equation}\label{shift_single}
    \Delta Q_{sc}=-\frac{r_pZ^2N_I}{2\pi\beta^2\gamma^3A\epsilon_i}.
\end{equation}
In case of the stacked beam, the emittance $\epsilon_i$ in the upper equation should be replaced by
\begin{equation}
	\frac{1}{2}\left(\epsilon_x+\sqrt{\epsilon_x\epsilon_y\tfrac{Q_x}{Q_y}}\right)
\end{equation}
and multiplied with the form factor $g_f$ to consider the non-homogeneous density profile of the stacked beam. For a homogeneous profile $g_f$ is $1$ and for a Gaussian distribution $g_f$ is $2$. $\epsilon$ is four times the rms emittance of the stacked beam. Suppose the vertical emittance remains constant ($\epsilon_{y}=\epsilon_i$) than the horizontal space charge tune shift for the stacked beam is
\begin{equation}\label{shift_multi}
    \Delta Q_{sc}^s=\Delta Q_{sc}\frac{2g_f}{d+\sqrt{\tfrac{d}{n_{eff}}\tfrac{Q_x}{Q_y}}}.
\end{equation}
During multi-turn-injection, the increase of the intensity and of the emittance usually compensate each other 
and the horizontal space charge tune shift remains below the initial one  
\begin{equation}
	\Delta Q_{sc}^s \lessapprox \Delta Q_{sc}.
\end{equation}
For the SIS18 this condition is fulfilled in all our simulation examples.
Other injection schemes, e.g. employing two planes with a tilted septum, could possibly lead to larger space charge parameters for the the stacked beam than for the incoming beamlet.
\begin{figure}
		\centering
		\includegraphics[scale=0.275]{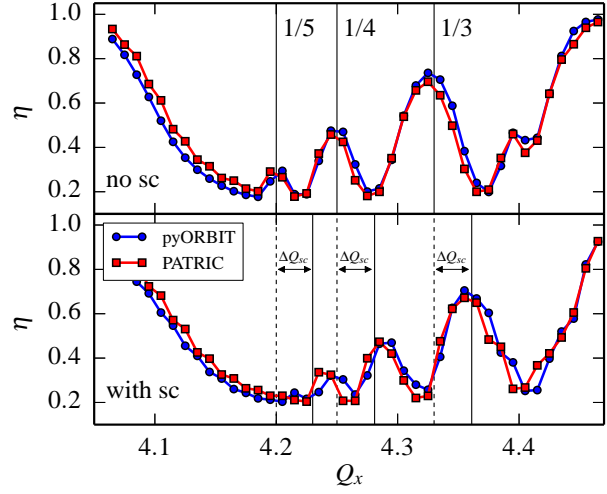}
\caption{Simulation result obtained with PyORBIT and PATRIC for the MTI efficiency as a function of the horizontal tune with and without space charge effects.}
\label{tune}       
\end{figure}
\\
Fig. \ref{tune} shows the effect of space charge on the MTI efficiency. One can observe that the maxima and minima of the MTI efficiency are shifted to the right by $\Delta Q_{sc}$ under the influence of space charge. The observed shift of the maxima/minima can be described very well through 
\begin{equation}\label{shift_n}
	(Q_x + \left|\Delta Q_{sc}\right|) n =integer.
\end{equation}
In the simulations the space charge tune shift for an injected beamlet was $\Delta Q_{sc}\approx-0.04$. \\
The simulations indicate that the presently chosen high working point $Q_x=4.17$ in the SIS18 is a good candidate in order to reach the FAIR design intensities. We do expect that the MTI efficiency for high beam currents (with space charge) will be close to the one achieved for low currents (low space charge) shown in Fig.~\ref{fig:exp}.
\begin{figure}
\centering
\includegraphics[scale=0.2]{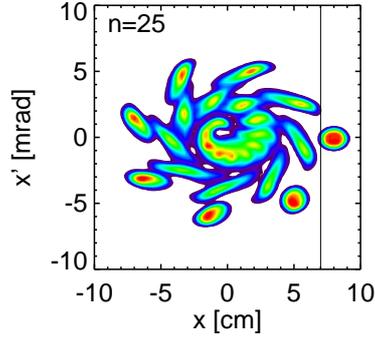}
\caption{The snapshot of an MTI simulation with space charge (same simulation parameters as in Fig.~\ref{fig:ps_nosc})}
\label{ps_sp}       
\end{figure}
\\
During and right after the injection process the stacked beam distribution will be non-stationary. 
The space charge potential energy will be larger than the potential energy of the corresponding matched equilibrium distribution. Therefore we can expect that the nonlinear space charge force will drive the beam distribution closer towards an equilibrium~\cite{Reiser:book}. 
In addition also chromaticity or beam mismatch causes a filamentation of the distribution between the injected beamlets. Fig. \ref{ps_sp} shows a snapshot of the broadening of the beamlets and a filamentation driven by space charge during the multi-turn injection. Besides the much higher intensity, the simulation parameters are similar to the ones chosen for Fig. \ref{fig:ps_nosc}. Especially close to the center, with space charge, the individual beamlets cannot be distinguished. At a latter stage a phase space distribution with a few larger holes can develop. In the center of Fig. \ref{ps_sp} the initial stage of such a hole can already be observed. For the chosen beam intensities the evolution of the beam distribution is dominated by the effect of space charge. Other effects, like the decoherence due to chromaticity, can be neglected.

\section{The role of the vertical emittance}
\begin{figure}[htb]
\centering	
\includegraphics[scale=0.25]{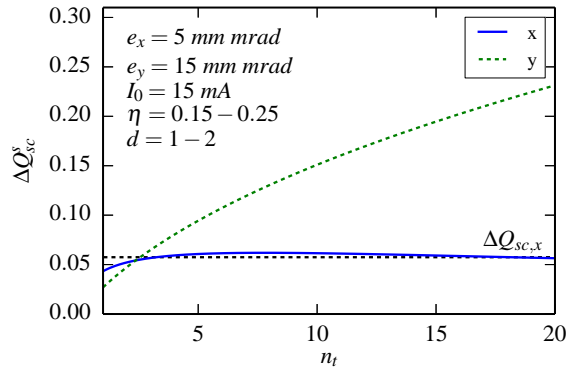}
\caption{Increase of the space charge tune shifts during injection if the vertical emittance remains constant. The vertical line indicates the horizontal space charge tune shift of the incoming beamlet.  \label{fig:mti} }
\end{figure}
In our simulation model the vertical emittance remains constant during and also after the horizontal MTI. The corresponding evolution of the vertical and horizontal space charge tune shifts are shown in Fig.~\ref{fig:mti} for the example case of an initial vertical emittance of $\epsilon_y=15$~mm~mrad. During the injection the beam loss as well as the dilution varies turn by turn. Presently the vertical emittance in the SIS18 increases after injection by $\frac{\Delta \epsilon_y}{\epsilon_y} \approx 0.2$. In \cite{paret:2010} the vertical emittance was obtained from the vertical beam profiles (for an $^{40}Ar^{18+}$ beam). The initial UNILAC emittance was measured by~\cite{Hayek2013} as $\approx 15$ mm mrad. Analytical estimates indicate that the emittance growth in the vertical plane is caused by injection steering errors and optical mismatch, as we will show in this section.  
The growth of the vertical emittance affects the horizontal space charge tune shift. More importantly the vertical emittance   determines the (vertical) space charge limit at $\Delta Q_y\approx -0.5$ in the SIS18, which is beyond the scope of the present study. In an optimal scenario the vertical emittance should increase in a controlled way up to the vertical machine acceptance, which is at $A_y=50$~mm~mrad in the SIS18 \cite{Spiller2008}. \\
During the horizontal MTI the vertical emittance increases in a non-controlled way due to steering errors and optical mismatch~\cite{Mohl2005,Fischer1996}. The emittance increases due to a radial displacement of the beam center after an injection steering error 
\begin{equation}
	\frac{\Delta \epsilon_y}{\epsilon_y} =\frac{\Delta r^2}{2\epsilon_y\beta_y},
\end{equation}
if the radial displacement is
\begin{equation}
	\Delta r = \sqrt{\Delta y^2+(\alpha_y\Delta y + \beta_y \Delta y')^2}.
\end{equation}
Assuming the steering errors are $\Delta y=0.5$ mm and $\Delta y'=0.5$ mrad the emittance growth is $\frac{\Delta \epsilon_y}{\epsilon_y} = 0.1$. The emittance growth is very sensitive to the slope error due to the multiplication with the beta function. A doubling of the slope error leads to $\frac{\Delta \epsilon_y}{\epsilon_y} = 0.4$.
For a mismatch of the lattice functions the emittance growth is
\begin{equation}
  \frac{\Delta \epsilon_y}{\epsilon_y} = F-1  
\end{equation}
with
\begin{equation}
F = \frac{1}{2}\left[\frac{\beta_0}{\beta_i}+\frac{\beta_i}{\beta_0}+\left(\alpha_0-\alpha_i\frac{\beta_0}{\beta_i}\right)^2\frac{\beta_i}{\beta_0}\right].
\end{equation}
Assuming the lattice functions are known with an error of 20$\%$ we obtain an emittance growth of $\frac{\Delta \epsilon_y}{\epsilon_y} = 0.2$. \\
A mismatch in the dispersion gives the following change in emittance
\begin{equation}
	\frac{\Delta \epsilon_y}{\epsilon_y} = \frac{\Delta D^2}{2\epsilon_y\beta_y}\sigma_p^2
\end{equation}
if the radial dispersion displacement is
\begin{equation}
	\Delta D = \sqrt{\Delta D_y^2+(\alpha_y\Delta D_y + \beta_y \Delta D_y')^2}.
\end{equation}
Since the injection kicker bends the incoming beamlet vertically there is a systematic vertical dispersion mismatch which results in $\frac{\Delta \epsilon_y}{\epsilon_y} = 5\times10^{-5}$.  \\
The total vertical emittance increase after injection due to the aforementioned effects should not be larger than $\frac{\Delta \epsilon_y}{\epsilon_y} = 0.3$, which is larger than the observed $\frac{\Delta \epsilon_y}{\epsilon_y}\approx 0.2$.

\section{Conclusions and Outlook} 

We studied the effect of transverse space charge on the multi-turn injection (MTI) using the example of the SIS18 at GSI. Two different simulation codes, PATRIC and pyORBIT, where adapted to the problem. The MTI quality (intensity and beam loss) depends very strongly on the horizontal tune. The beam loss maxima are located at the lower order resonances. An optimal fractional tune is identified at 1/6. The transverse space charge force causes a shift of the beam loss maxima and of the optimum tune.  We find that this shift corresponds to the space charge tune shift of the incoming beamlets. For the SIS18 conditions we find an exponential decrease of the orbit bump results in good compromise between compact stacking and beam loss. More or less independently from the injected emittance the beam loss could be reduced to $\eta=0.15$ due to a rather compact packing (e.g. $d=2$). A direct comparison of the beam distribution with and without space charge indicates that the nonlinear space charge  force between the injected beamlets drives the final beam distribution towards a more homogeneous density with a lower total potential energy. Future studies will focus on the application of the skew quadrupoles in the SIS18 to possibly increase the number of injected turns through a coupling with the vertical plane. Another option to increase the MTI efficiency is to ramp the tune during injection. 

\section*{Acknowledgement}
The authors thank Y. El-Hayek for his measurement results and  D. Ondreka for useful discussions about the injection into the SIS18.

\section*{References}

\bibliographystyle{elsarticle-num}
\bibliography{mti_sis18}

\end{document}